# Kepler's Constant and WDS Orbit

S. Siregar[1]* and R. N. Pratama

[1]*Bosscha Observatory and Astronomy Research Division, Faculty of Math. and Nat. Sci. – ITB, Bandung, Indonesia*
*E-mail: suryadi@as.itb.ac.id

**ABSTRACT**
The aim of this work are to find a Kepler's constant by using polynomial regression of the angular separation $\rho = \rho(t)$ and the position angle $\theta = \theta(t)$. The Kepler's constant obtained is used to derive the elements of orbit. As a case study the angular separation and the position angle of the WDS 00063 +5826 and the WDS 04403-5857 were investigated. For calculating the element of orbit the Thiele-Innes van den Bos method is used. The rough data of the angular separation $\rho(t)$ and the position angle $\theta(t)$ are taken from the US Naval Observatory, Washington. This work also presents the masses and absolute bolometric magnitudes of each star. These stars include into the main-sequence stars with the spectral class G5V for WDS 04403-5857 and the type of spectrum G3V for WDS 00063 +5826. The life time of the primary star and the secondary star of WDS 04403-5857 nearly equal to $2\times10^{10}$ years. The life time of the primary star and the secondary star of WDS 00063+5826 are $2\times10^{10}$ years and $1.9\times10^{10}$ years, respectively.

## 1   INTRODUCTION

Since 1928 the Bosscha Observatory is engaged in visual double stars studies, with the main purpose to develop new observational techniques in order to obtain results with accuracy greater than that attained till now. Two different observing techniques in the last decade have been developed for double star researches: the photographic multiple exposures method and the CCD measurements. On the other hands theoretical studies on the wide double stars are also in progress. The importance of visual double stars to astronomy needs to be emphasized. A study of their orbit permits the calculation of the sum of the masses of the components if we also have the parallax of the system. The first stage of orbit computation consists of the calculation of the apparent orbit of the secondary with respect to the primary, usually the brighter of the two component. A measurement of visual double stars comprises two components: the angular separation between the components, $\rho$, the position angle $\theta$ of the secondary with respect to the primary, as function of the time observation. According to two-body problem the angular momentum of this system is constant given by the second Kepler's law:

$$C = \rho^2 \frac{d\theta}{dt} = \text{constant} \qquad (1)$$

In this investigation we utilized the data from the Washington Double Star Catalog (http://ad.usno.navy.mil/wds) that provides all of the information of the position angle $\theta$ and the angular separation $\rho$.

To determine the orbits of visual double stars there are several methods that can be used, for examples, the method of Thiele-Innes-van den Bos, the method of Kowalsky, and the method of Glassnap. In this work visual double star orbits will be sought by the method of Thiele-Innes van den Bos. In this case the Kepler's constant $C$ plays an important role. Changes in the value of $C$ will lead to a fairly significant error in determining the orbital elements. Therefore Kepler's constant $C$ must be determined accurately before deriving orbital elements.

## 2   METHODOLOGY

### 2.1   Calculating the Kepler's Constant

Data used in this work for WDS 00063 +5826 starts with the data taken in 1823 and the last data is from 2010 while for WDS 04403-5857 the first data taken in 1835 and last in 2002. To discover the constant $C$, we use the following algorithm.

1. Determine the polynomial regression of $\rho = \rho(t)$ and $\theta = \theta(t)$. By taking the assumption of the current measurement technique is better than the past, then for the last data given greater weight than previous data.
2. Based on polynomial regression derived by the method of "Least Square" Kepler's constant $C$, and standard deviation can be calculated. If the standard deviation is relatively small the process is stopped, otherwise proceed to step 3.
3. By using the constants $C$ and take the time interval $\Delta t$ find the new value of $\rho = \rho(t)$, and $\theta = \theta(t)$.
4. Calculate the new Kepler's constant $C$, and compare with the previous value. And if the test

$|C_{i+1} - C_i| \le \varepsilon$ is satisfied the process terminates, otherwise return to step 3 by taking the value of $\Delta t$ is smaller.

Figure 1, 2, 3 and 4 presents the data of the angular separation and the position angle of WDS 04403-5857 and WDS 00063 +5826.

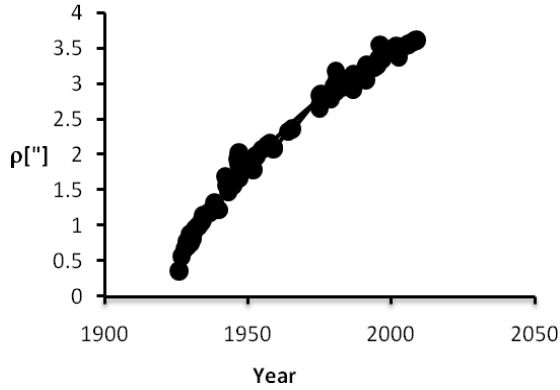

**Figure 1**. The angular separation of WDS 04403-5857 as function of the epoch of observation.

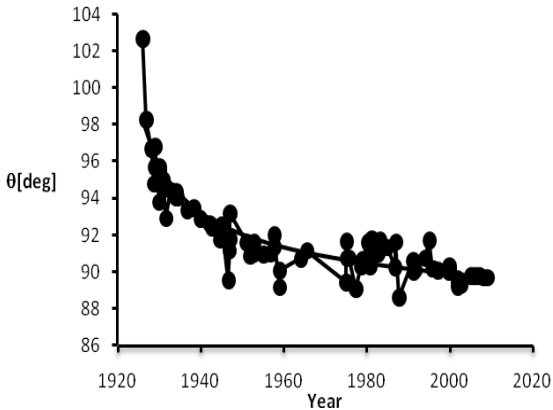

**Figure 2**. The position angle of WDS 04403 -5857 as function of the epoch of observation.

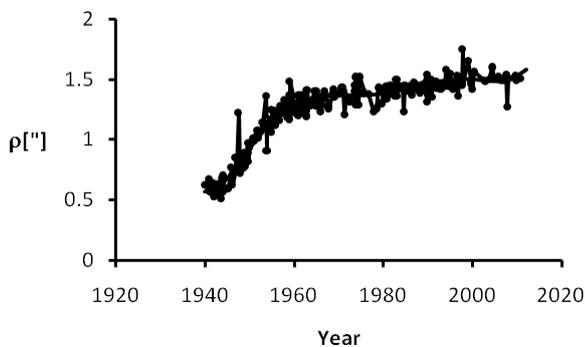

**Figure 3**. The angular separation of WDS 00063 +5826 as function of the epoch of observation.

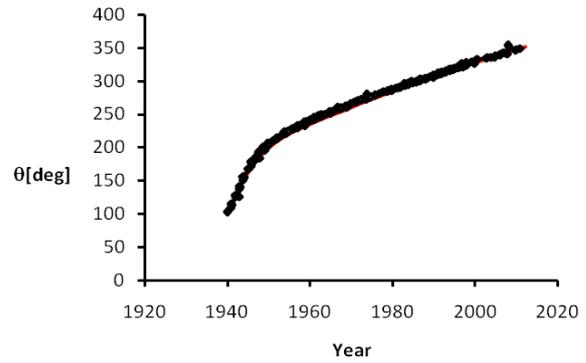

**Figure 4**. The position angle of WDS 00063+5826 as function of the epoch of observation.

Kepler's constant obtained for two pairs of visual double stars WDS 04403-5857 and WDS 00063 +5826 are presented in Table 1.

**Table 1**. Kepler's constant C, for the pair WDS 04403-5857 and WDS 00063 +5826.

| WDS 04403-5857 Epoch 2000 | | WDS 00063 +5826 Epoch 2000 | |
|---|---|---|---|
| $C$ | $\sigma$ | $C$ | $\sigma$ |
| 0.005 | .00000017 | 0.079 | .00000017 |

The symbols $C$ and $\sigma$ represents Kepler's constant and standard deviation in [rad arcsec$^2$yr$^{-1}$], respectively.

### 2.2 Thiele-Innes van den Bos Method

Because the procedure is relatively easy, many people have used the method of Thiele-Innes van den Bos for calculating the orbits of visual double stars. We have used this method to calculate some of the double star orbits. For example, that conducted by Siregar (1988), Siregar & Hadi Nugroho (2006), Siregar & Kuncarayakti (2007) and Siregar (2008). In this method we define the following notation.

The symbols: $P$, $T$, $e$, $a$, $i$, $\omega$, $\Omega$ have their usual meaning for orbits; $\mu = 2\pi/P$ mean annual motion, $t_i$ = time at which the secondary star occupies its i-th position, $E_i$ = anomaly eccentric at time $t_i$.

$$t_{ij} = t_j - t_i \qquad (2)$$

$$E_{ij} = E_j - E_i \qquad (3)$$

$$V = E_{12};\ U = E_{23}\ \text{and}\ W = E_{12} \qquad (4)$$

$\theta_i$, $\rho_i$: coordinates of the secondary star at time $t_i$.

$$\varDelta_{ij} = \rho_i\, \rho_j\, sin(\theta_j - \theta_i) \qquad (5)$$

$$E_{ij} - \sin E_{ij} = \mu\left(t_{ij} - \frac{\Delta_{ij}}{C}\right) \quad (6)$$

From the equation (6) the value of $\mu$ calculated by using the method of the trial & error by condition $|E_{31} - (E_{32} + E_{21})| \leq \varepsilon$, the symbol $\varepsilon$ is the relative precession. Furthermore to determine the orbital elements we are able to transform the angular separation $\rho(t)$ and the position angle $\theta(t)$ to the orbital elements. For more details explanations see Alzner (2003).

## 3 RESULTS AND DISCUSSION

### 3.1 Element of Orbit

The elements of orbit calculated by Thiele Innes van den Bos method are presented in Table 2.

**Table 2**. The elements of orbit of the test stars calculated by Kepler's constant from Table 1.

| No | Element | WDS 04403-5857 | WDS 00063+5826 |
|---|---|---|---|
| 1. | $P$[yrs] | 360.36 | 315 |
| 2. | $T_o$ | 1932.74 | 2200.45 |
| 3. | $e$ | 0.67 | 0.52 |
| 4. | $a$ | 3″.051 | 2″.572″ |
| 5. | $i$ | 90°.094 | 92°.184 |
| 6. | $\Omega$ | 87°.42 | 88°.97 |
| 7. | $\omega$ | 264°.98 | 177°.46 |
| 8. | $p$ | 0″.053 | 0″.046 |

### 3.2 Mass and Absolute Bolometric Magnitude

The apparent magnitude of the test stars are taken from the U.S. Naval Observatory, Washington. Primary and secondary of WDS 04403-5857 (type G5V spectrum) are, respectively, 7.33 and 7.45 mag, while those for WDS 00063 +5826 (G3V) are 6.42 and 7.32 mag. To obtain apparent bolometric magnitude we used the bolometric correction proposed by Flower (1996), and by using (7), (8) and (9) simultaneously. The masses and the absolute bolometric magnitudes of the stars can be derived by iteration. The results are presented in Table 3.

$$p = \frac{a}{\sqrt[3]{P^2(M_1 + M_2)}} \quad (7)$$

$$\log(M/M_\odot) = 0.1(4.6 - M_{bol}), \quad 0 \leq M_{bol} \leq 7.5 \quad (8)$$

$$\log(M/M_\odot) = 0.145(4.6 - M_{bol}),$$
$$7.5 \leq M_{bol} \leq 11 \quad (9)$$

**Table 3**. The masses and the absolute bolometric magnitudes of WDS 04403-5857 and WDS 00063 +5826.

| No | Element | WDS 04403-5857 | WDS 00063+5826 |
|---|---|---|---|
| 1. | $^1M_{bol}$ | 5.88 | 5.68 |
| 2. | $^2M_{bol}$ | 6.00 | 5.80 |
| 3. | $M_1$ | $0.774 M_\odot$ | $0.779 M_\odot$ |
| 4. | $M_2$ | $0.724 M_\odot$ | $0.758 M_\odot$ |

These stars belong to the main sequence stars. If the life time of the Sun is $t_0$ and $M_\odot$ is mass of the Sun, the life time of the star can be determined from the well-known relation:

$$t = t_0 (M/M_\odot)^{-2.5} \quad (10)$$

By considering $t_0 \approx 10^{10}$ the life time of the primary star and the secondary star of WDS 04403-5857 nearly equal to $2\times 10^{10}$ years, almost twice longer. For the system of WDS 00063+5826 the life time of the primary star and the secondary star of WDS 00063+5826 are $2\times 10^{10}$ years and $1.9\times 10^{10}$ years, respectively.

## 4 CONCLUSION

The quality of the orbits obtained in this way still needs to be fixed. Developing other methods of determining the Kepler's constant is needed. The future researches on wide double stars in our Observatory can be summarized as follows:
1. Continuing the observing programme of wide binary systems obtained by means of new techniques using CCD detectors.
2. Studying the dynamics of wide binary systems by means of using the available catalogues, parallaxes, and radial velocities.
3. Performing statistical work concerning with the systematic and accidental errors detected in visual double star observations, obtained by the Zeiss-60cm double refractor, with other techniques over sample of well-known orbits.

### Acklowledgement